\PassOptionsToPackage{unicode}{hyperref}
\PassOptionsToPackage{hyphens}{url}
\documentclass[
]{article}
\usepackage{amsmath,amssymb}
\usepackage{iftex}
\ifPDFTeX
  \usepackage[T1]{fontenc}
  \usepackage[utf8]{inputenc}
  \usepackage{textcomp} 
  \usepackage{lmodern}
\else 
  \usepackage{unicode-math} 
  \defaultfontfeatures{Scale=MatchLowercase}
  \defaultfontfeatures[\rmfamily]{Ligatures=TeX,Scale=1}
\fi
\IfFileExists{upquote.sty}{\usepackage{upquote}}{}
\IfFileExists{microtype.sty}{
  \usepackage[]{microtype}
  \UseMicrotypeSet[protrusion]{basicmath} 
}{}
\makeatletter
\@ifundefined{KOMAClassName}{
  \IfFileExists{parskip.sty}{%
    \usepackage{parskip}
  }{
    \setlength{\parindent}{0pt}
    \setlength{\parskip}{6pt plus 2pt minus 1pt}}
}{
  \KOMAoptions{parskip=half}}
\makeatother
\usepackage{xcolor}
\usepackage{longtable,booktabs,array}
\usepackage{multirow}
\usepackage{calc} 
\usepackage{etoolbox}
\makeatletter
\patchcmd\longtable{\par}{\if@noskipsec\mbox{}\fi\par}{}{}
\makeatother
\IfFileExists{footnotehyper.sty}{\usepackage{footnotehyper}}{\usepackage{footnote}}
\makesavenoteenv{longtable}
\usepackage{graphicx}
\makeatletter
\def\maxwidth{\ifdim\Gin@nat@width>\linewidth\linewidth\else\Gin@nat@width\fi}
\def\maxheight{\ifdim\Gin@nat@height>\textheight\textheight\else\Gin@nat@height\fi}
\makeatother
\setkeys{Gin}{width=\maxwidth,height=\maxheight,keepaspectratio}
\makeatletter
\def\fps@figure{htbp}
\makeatother
\ifLuaTeX
  \usepackage{selnolig}  
\fi
\IfFileExists{bookmark.sty}{\usepackage{bookmark}}{\usepackage{hyperref}}
\IfFileExists{xurl.sty}{\usepackage{xurl}}{} 
\hypersetup{
  hidelinks,
  pdfcreator={LaTeX via pandoc}}

\author{}
\date{}

\begin{document}

\textbf{A State--Transition Information Space for Time-Series Dynamics: Theory and Application}

Dragutin T Mihailović\textsuperscript{a,b}, Vijay P. Singh\textsuperscript{c}, and Slavica Malinović-Milićević\textsuperscript{d,e}

\textsuperscript{a}Department of Physics, Faculty of Sciences, University of Novi Sad, Trg Dositeja Obradovića 4, 21000 Novi Sad, Serbia; \href{mailto:guto@df.uns.ac.rs}{\nolinkurl{guto@df.uns.ac.rs}}

\textsuperscript{b}Department of Atmospheric and Environmental Sciences, University at Albany, State University of New York, Albany, NY 12222, USA

\textsuperscript{c}Department of Biological and Agricultural Engineering and Zachry Department of Civil \& Environmental Engineering, Texas A\&M University, College Station, TX 77843-2117, USA; vsingh@tamu.edu

\textsuperscript{d}Geographical Institute "Jovan Cvijić" SASA, Đure Jakšića 9,11000 Belgrade, Serbia; s.malinovic-milicevic@gi.sanu.ac.rs

\textsuperscript{e}Institute of Environmental Engineering, Peoples\textquotesingle{} Friendship University of Russia (RUDN University), 6 Miklukho-Maklaya St, 117198 Moscow, Russian Federation

Author for correspondence: Dragutin T Mihailović, e-mail: \href{mailto:guto@df.uns.ac.rs}{\nolinkurl{guto@df.uns.ac.rs}}

\textbf{ORCIDs}: DTM 0000-0002-8380-1844, VPS 0000-0003-1299-1457, SM-M 0000-0001-9696-6982

\textbf{Abstract}

Characterizing dynamical organization in time series requires distinguishing the diversity of accessible states from uncertainty in their temporal transitions. Here we introduce a state--transition information space based on two normalized entropy measures derived from ordinal patterns: \(K_{q\ }\), quantifying ordinal-state diversity, and \(K_{t}\), quantifying transition uncertainty. Their joint \(K_{t} - K_{q}\ \)representation provides a two-dimensional framework in which dynamical regimes and their temporal evolution can be examined. Theoretical properties establish the bounded relation\(\ 0 \leq K_{t} \leq K_{q} \leq 1\), while canonical time series identify distinct empirical domains ranging from ordered dynamics to near-maximal randomness. Sliding-window analysis further shows that systems can exhibit temporal trajectories through the \(K_{t} - K_{q}\) plane rather than remaining at a fixed dynamical state. Application to 1,879 monthly naturalized streamflow records from the U.S. rivers shows that river dynamics occupy a distinct intermediate region between ordered and highly disordered regimes. Moreover, river positions shift systematically toward higher \(K_{t}\) and \(K_{t}\) with increasing drainage area, revealing scale-dependent organization of streamflow dynamics. The framework therefore provides a compact means of comparing state diversity, transition uncertainty, and their temporal and spatial organization across time-series systems.


\newpage
\textbf{1. Introduction}

Complex systems arise throughout nature, engineering, geophysics, meteorology, ecosystems, and social and economic sciences. Their behaviour emerges from interactions among many components, producing dynamics that cannot be inferred directly from the properties of individual elements. Examples include turbulent flows, climate variability, neural activity, ecological networks, financial markets, and river discharge. Despite their diversity, these systems share a common challenge: developing quantitative representations that distinguish deterministic organization from stochastic variability while revealing the mechanisms governing temporal evolution.

Classical phase-space analysis characterizes dynamics using physical state variables and deterministic equations of motion. Although highly successful for model-based systems, such\\
representations are often unavailable for empirical observations, where only scalar time series are accessible, and the governing equations are unknown. Information-theoretic approaches provide an attractive alternative by quantifying dynamical properties directly from observations. Shannon entropy measures statistical uncertainty {[}1{]}, while information theory extends this concept to stochastic processes and entropy rates that characterize information production associated with state transitions {[}2{]}. Consequently, numerous entropy-based measures, including permutation entropy, sample entropy, multiscale entropy, and related approaches, have advanced the characterization of nonlinear dynamics across the physical, biological, and environmental sciences.

Information-based geometric representations provide a natural extension of these scalar measures. The ordinal-pattern framework introduced by {[}3{]} provides a robust symbolic representation of time series, while the entropy--complexity plane developed by {[}4{]} combined normalized permutation entropy with a statistical complexity measure derived from the ordinal-pattern probability distribution. This approach demonstrates that periodic, chaotic, and stochastic processes can occupy distinct regions of a low-dimensional entropy--complexity plane.

Subsequent developments, including the complexity--entropy causality plane {[}5{]} and advances in ordinal symbolic analysis {[}6{]}, have further established information-geometric methods as effective tools for investigating nonlinear and empirical data.

More recently, the field has been moving rapidly beyond static ordinal-state distributions toward approaches that explicitly incorporate temporal transitions, transition entropy, and network organization. This emerging transition-based direction has extended ordinal analysis to applications ranging from the reconstruction of chaotic signals {[}7{]} and the identification of resilience changes in complex systems {[}8{]} to graph-based representations of ordinal information {[}9{]}. In parallel, permutation transition entropy has incorporated Markov transitions between ordinal patterns {[}10{]}, while transition-based complexity--entropy diagrams have extended conventional entropy--complexity representations by incorporating transition probabilities and conditional entropy {[}11{]}. Related approaches based on ordinal transition networks have further demonstrated that transition information can provide valuable insight into synchronization, connectivity, and statistical complexity in nonlinear systems {[}12-14{]}. These five closely related studies {[}10-14{]} bridge classical ordinal-pattern analysis with transition-based representations of complex dynamics, shifting the focus from the occurrence of symbolic states to their temporal organization and evolution.

Despite these advances, conventional entropy--complexity representations remain primarily distribution-based. They characterize the diversity and distribution of accessible symbolic states, but do not explicitly quantify how these states are connected through time. Structural complexity and temporal complexity therefore represent distinct aspects of dynamical organization that are not explicitly separated within conventional entropy--complexity frameworks. Characterizing the dynamical organization in time series consequently requires distinguishing the diversity of accessible states from the uncertainty associated with their temporal transitions.

Motivated by this distinction and the developing transition-based perspective, we introduce a two-dimensional state--transition information space, based on two normalized entropy measures derived from ordinal patterns: \(K_{q}\), quantifying ordinal-state diversity, and \(K_{t}\ \), quantifying uncertainty in transitions between successive ordinal states. Based on ordinal symbolic dynamics and the corresponding transition probability matrix, the framework integrates state occupancy and transition organization within a unified information-theoretic description. Their joint representation in the \(K_{t} - K_{q}\ \)plane provides a framework in which different dynamical regimes and their temporal evolution can be examined.

Unlike conventional entropy--complexity planes, in which a time series is generally represented by a single point, the proposed framework represents each moving window of a time series by a point in the \(K_{t} - K_{q}\) plane. Successive points therefore form a trajectory, allowing changes in dynamical organization and transitions between regimes to be examined over time.

The framework reveals a geometric organization of deterministic, chaotic, stochastic, and empirical systems in the \(K_{t} - K_{q}\ \)plane. It also establishes finite-sample constraints that define the admissible information space and provides improved discrimination between systems with similar structural complexity but different temporal complexity. By jointly representing state diversity and transition uncertainty, the proposed information space provides a complementary background for analyzing complex time series and characterizing their temporal organization.

The remainder of the paper is organized as follows. Section 2 presents the main theorems, their proofs, and the finite-sample constraints of the information space. Section 3 applies the\\
framework to canonical deterministic, chaotic, and stochastic systems to establish its geometric organization and interpretive domains. Section 4 examines empirical time series and demonstrates the ability of the framework to distinguish dynamical regimes and temporal organization. Section 5 discusses the main findings, limitations, and potential applications of the proposed approach. Finally, Section 6 is reserved for concluding remarks.

\textbf{2. Theoretical Background}

\textbf{2.1 Preliminaries}

The complexity of a dynamical system may be characterized from several complementary perspectives. In the present framework, we distinguish between structural complexity and temporal complexity, which describe different aspects of the organization of a time series. Structural complexity refers specifically to the diversity and statistical distribution of the symbolic states accessible to the system, independently of the temporal order in which those states occur. It is therefore quantified by the entropy of the state distribution and, in the present framework, represented by the normalized state entropy \(K_{q}\).

This use of the term structural complexity should be distinguished from the statistical complexity introduced in computational mechanics by {[}15{]}. The latter quantifies the information stored in the minimal predictive representation of a dynamical system, expressed through its causal states. Although both concepts address the organization of dynamical systems, they characterize fundamentally different properties: \(K_{q}\) measures the statistical diversity and distribution of observed symbolic states, whereas computational statistical complexity measures the information required to represent the system\textquotesingle s predictive causal structure.

By contrast, temporal complexity characterizes the organization and uncertainty of transitions between successive states. It concerns not only which states are visited, but also how the system evolves from one state to the next. In the present framework, temporal complexity is quantified by the conditional entropy of the next symbolic state given the current state, and its normalized form is denoted by \(K_{t}\). Thus, \(K_{q}\) describes the complexity associated with state diversity and distribution, whereas \(K_{t}\) describes the complexity associated with temporal transitions and conditional uncertainty.

This distinction provides the conceptual basis for representing time-series dynamics in the two-dimensional \(K_{t} - K_{q}\) state--transition information space.

Structural and temporal complexity are therefore not necessarily correlated. A system may visit a large number of distinct states with a broad or nearly uniform state distribution, corresponding to high structural complexity, while its transitions remain highly constrained and predictable, corresponding to low temporal complexity. Conversely, a system may exhibit a diverse state distribution together with highly uncertain transitions, resulting in both high structural and high temporal complexity.

\textbf{2.2 Basic Theorems}

The mathematical formulation developed in this section builds on three established foundations. The ordinal-state representation of the time series follows the framework introduced by {[}3{]}. The definitions and properties of Shannon entropy, conditional entropy, and mutual information are based on the information-theoretic framework of {[}1{]} and its formal treatment by {[}2{]}. The representation of the symbolic dynamics in terms of state probabilities and transition matrices, together with the first-order Markov property, follows the standard theory of finite-state Markov chains {[}16{]}. These established results provide the mathematical basis for the state--transition representation and the theoretical results developed below. Further mathematical details and related properties are provided in Supplementary Sections S1.1--S1.6.

\textbf{Theorem 1. State-Transition Representation}

Let \{\(S_{t}\)\} be a stationary ordinal-state process with finite state space

\(\Omega = \left\{ S_{1},\ldots,S_{A}\  \right\}\),

where \(A = d!\ \)is the total number of possible ordinal states (permutations) for an embedding dimension \(d\). Let

\(\pi_{i} = \ P\left( S_{t} = S_{i} \right)\)

denote the stationary state probabilities, and let

\(P_{ij} = \ P(S_{t + 1} = \ S_{j}\ |\ S_{t} = \ S_{i}\))

denote the transition probabilities. Then, the joint probability of two consecutive ordinal states is

\[P(S_{t} = S_{i},S_{t + 1} = S_{j}) = \pi_{i}P_{ij}.\]

If, in addition, \{\(S_{t}\)\} is a first-order Markov process, then the pair \((\pi,P)\) uniquely determines all finite-dimensional joint distributions of the process. In particular,

\[P\left( S_{1} = S_{i_{1}},\ldots,S_{m}{= S}_{i_{m}} \right) = \pi_{i_{1}}P_{i_{1\ }i_{2}}{P_{i_{2\ }i_{3}}P}_{i_{m - 1\ }i_{m}}.\]

\textbf{Proof.}

By the definition of conditional probability,

\[P(S_{t} = S_{i},S_{t + 1} = S_{j}) = \pi_{i}P_{ij}.\]

Thus, the pair (\(\pi\), P) determines the joint probability distribution of two consecutive ordinal states.

Now assume, in addition, that \{\(S_{t}\)\} is a first-order Markov process. For any sequence of \(m\ \)ordinal states, the chain rule of probability gives

\[P\left( S_{1} = S_{i_{1}},\ldots,S_{m}{= S}_{i_{m}} \right) = P\left( S_{1} = S_{i_{1}} \right)\prod_{k = 2}^{m}{P\left( S_{k} = S_{i_{k}}\mid S_{1} = S_{i_{1}},\ldots,S_{k - 1}{= S}_{i_{k - 1}} \right)}.\]

Because the process is a first-order Markov,

\[P\left( S_{k} = S_{i_{k}}\mid S_{1} = S_{i_{1}},\ldots,S_{k - 1}{= S}_{i_{k - 1}} \right) = P\left( S_{k} = S_{i_{k}}\mid S_{k - 1} = S_{i_{k - 1}} \right).\]

Furthermore,

\[P\left( S_{1} = S_{i_{1}} \right) = \pi_{i_{1}}.\]

Therefore,

\[P\left( S_{1} = S_{i_{1}},\ldots,S_{m}{= S}_{i_{m}} \right) = \pi_{i_{1}}\prod_{k = 2}^{m}{P_{i_{k - 1\ }}P_{i_{k\ }}}.\]

Hence, once the stationary state probabilities \(\pi\) and the transition matrix \(P\) are specified, every finite-dimensional joint distribution of the stationary first-order Markov ordinal-state process is uniquely determined. Thus, the pair \((\pi,P)\) provides a complete probabilistic representation of the process under the first-order Markov assumption. The relation of this representation to first-order Markov chains and one-step entropy rates is further discussed in Supplementary Section S1.6. \(\blacksquare\)

\textbf{Theorem 2. Bounds of the} \(\mathbf{K}_{\mathbf{t}}\mathbf{-}\mathbf{K}_{\mathbf{q}}\) \textbf{Phase Space}

For a stationary finite-state ordinal process, define the normalized structural and transition entropies as

\[K_{q} = \frac{H\left( S_{t} \right)}{\log A},\]

\[K_{t} = \frac{H\left( S_{t + 1}\mid S_{t} \right)}{\log A}.\]

Then

\[{{0 \leq K}_{t} \leq K}_{q} \leq 1.\]

\textbf{Proof.}

The non-negativity of conditional entropy gives

\(H\left( S_{t + 1}\mid S_{t} \right) \geq 0\).

Moreover, conditional entropy cannot exceed the corresponding marginal entropy:

\[H\left( S_{t + 1}\mid S_{t} \right) \leq \ H\left( S_{t + 1} \right).\]

Since the process is stationary,

\[H\left( S_{t} \right) = H\left( S_{t + 1} \right) = H_{q}.\]

Because\({\ S}_{t}\) is a discrete random variable with \(A\) possible states,

\[H_{q} = H\left( S_{t} \right) \leq \log A.\]

Therefore,

\[{{0 \leq H}_{t} \leq H}_{q} \leq \log A,\]

where

\[H_{t} = H\left( S_{t + 1}\mid S_{t} \right).\]

The entropy bounds and their transition-matrix formulation, including the distinction between the stationary and nonstationary cases, are further developed in Supplementary Section S1.2.

Since \(\log A > 0\), division by \(\log A\) yields

\({{0 \leq K}_{t} \leq K}_{q} \leq 1\).

Therefore, the admissible region of the stationary \(K_{t} - K_{q}\ \)phase space is the triangular region bounded by \(K_{t} = 0\), \(K_{t} = K_{q}\), and\({\ K}_{q} = 1\).\(\ \)The uniform-transition limit \(K_{q} = K_{t} = 1\), corresponding to maximal state diversity and maximal one-step transition uncertainty, is derived in Supplementary Section S1.5.\(\blacksquare\)

\textbf{Theorem 3. Mutual-Information Interpretation of Phase-Space Separation}

For a stationary ordinal-state process,

\[{K_{q} - K}_{t} = \frac{{I(S}_{t};S_{t + 1})}{\log A},\]

where\(\ {I(S}_{t};S_{t + 1})\) denotes the mutual information between consecutive ordinal states.

\textbf{Proof.}

By the definition of mutual information,

\[{I(S}_{t};S_{t + 1}) = H\left( S_{t + 1} \right) - H\left( S_{t + 1}\mid S_{t} \right).\]

By stationarity,

\[H\left( S_{t} \right) = H\left( S_{t + 1} \right) = H_{q}\]

and, by definition,

\(H\left( S_{t + 1}\mid S_{t} \right) = H_{t}\).

Therefore,

\[{I(S}_{t};S_{t + 1}) = H_{q} - H_{t}.\]

Dividing by\(\ \log A\), we obtain

\[\frac{{I(S}_{t};S_{t + 1})}{\log A} = \frac{H_{q} - H_{t}}{\log A} = \frac{H_{q}}{\log A} - \ \frac{H_{t}}{\log A}.\]

Using the definitions of \(K_{q}\ \)and \(K_{t}\), it follows that

\[{K_{q} - K}_{t} = \frac{{I(S}_{t};S_{t + 1})}{\log A}.\]

Thus, the vertical separation between the coordinates \(K_{q}\) and \(K_{t}\) is exactly the mutual information between consecutive ordinal states normalized by the maximum entropy \(\log A\). The corresponding entropy decomposition and normalized mutual-information relation are developed further in Supplementary Section S1.1. \(\blacksquare\)

\textbf{Corollary 1. Interpretation of the Diagonal}

For a stationary ordinal-state process,

\({K_{q} = K}_{t} \Longleftrightarrow \ {I(S}_{t};S_{t + 1}) = 0\).

Thus, the diagonal

\[{K_{q} = K}_{t}\]

represents statistical independence between consecutive ordinal states.

\textbf{Proof.}

From Theorem 3,

\[{K_{q} - K}_{t} = \frac{{I(S}_{t};S_{t + 1})}{\log A}\]

Since \(\log A > 0\) ,

\[{K_{t} = K}_{q}\]

if and only if

\[{I(S}_{t};S_{t + 1}) = 0.\]

For discrete random variables, vanishing mutual information is equivalent to statistical independence. Therefore,

\[{K_{t} = K}_{q} \Leftrightarrow {I(S}_{t};S_{t + 1}),\]

if and only if \(S_{t}\ \)and \(S_{t + 1}\ \)are statistically independent. The equality condition and its equivalent transition-matrix representation are given in Supplementary Section S1.3. \(\blacksquare\)

\textbf{Corollary 2. Deterministic Transition Limit}

If the next ordinal state is uniquely determined by the current ordinal state, then

\[- \sum_{i}^{}\pi_{i}\sum_{j}^{}{P_{ij}\log P_{ij} = 0},\]

and consequently,

\[K_{t} = \frac{H_{t}}{\log A} = 0.\]

\textbf{Proof.}

If the next ordinal state is uniquely determined by the current ordinal state, then, for every state \(S_{i}\) with \(\pi_{i} > 0\), there exists a unique state \(S_{j}\) such that \(P_{ij} = 1\), while \(P_{ik} = 0\ \)for all \(k \neq j\).

Therefore,

\[- \sum_{i}^{}\pi_{i}\sum_{j}^{}{P_{ij}\log P_{ij} = 0}.\]

Hence,

\[H_{t} = - \sum_{i}^{}\pi_{i}\sum_{j}^{}{P_{ij}\log P_{ij}\log 1 = 0}.\]

Since \(\log A\) \textgreater{} 0,

\[K_{t} = \frac{H_{t}}{\log A} = 0.\]

Thus, deterministic transitions lie on the lower boundary \(K_{t}\)= 0 of the \({K_{t} - K}_{q}\)

phase space. The detailed deterministic-transition limit and its distinction from low structural entropy are given in Supplementary Section S1.4. \(\blacksquare\)

The theoretical formulation above provides the basis for empirical estimation from finite time-series records. The implementation of the framework for finite windows, the finite-sample criterion, sliding-window trajectories, and the interpretation of phase-space locations are described in the canonical-system analysis below, with additional technical details provided in Supplementary Sections S1.7--S1.17.

\textbf{2.3 Physical Interpretation of the} \(\mathbf{K}_{\mathbf{t}}\mathbf{-}\mathbf{K}_{\mathbf{q}}\mathbf{\ }\)\textbf{Phase Space}

The \(K_{t} - K_{q}\ \)phase space provides a two-dimensional description of temporal dynamics by separating \emph{state diversity} from \emph{one-step transition uncertainty}. The quantity\({\ K}_{q}\), defined as the normalized Shannon entropy of the ordinal-state distribution, characterizes state diversity\textbf{,} that is, the diversity of ordinal states explored by the system: high \(K_{q}\ \)indicates that a broad range of states is visited, whereas low \(K_{q}\ \)indicates that the dynamics are concentrated on a restricted subset of possible states. In contrast, \(K_{t}\) quantifies the uncertainty of the next state conditional on the present state. Low \(K_{t}\ \)therefore indicates strongly constrained transitions and greater one-step predictability, whereas high \(K_{t}\ \)indicates a broader set of possible successor states and greater transition uncertainty.

The distinction between these two quantities is essential, because state diversity and temporal organization are not equivalent properties. Two systems may have similar \(K_{q}\), and hence explore states with comparable overall diversity, while exhibiting substantially different \(K_{t}\) values because their transitions between those states are organized differently. Conversely, systems with similar \(K_{t}\ \)may have different \(K_{q}\)values if they possess comparable transition uncertainty but explore different numbers or distributions of states. Thus, \(K_{q}\ \)describes \emph{which states the system explores}, whereas \(K_{t}\ \)describes \emph{how predictably it moves between them}.

The difference between the two quantities is given by

\[{K_{q} - K}_{t} = \frac{{I(S}_{t};S_{t + 1})}{\log A},\]

where \({I(S}_{t};S_{t + 1}\ \)is the mutual information between successive ordinal states and \(A = d!\) is the number of possible ordinal states. Consequently, \({K_{q} - K}_{t}\ \)measures the normalized one-step temporal dependence. A large separation between \(K_{q}\ \)and \(K_{t}\ \)indicates that knowledge of the present state strongly constrains the next state, whereas when \(K_{t}\) is close to \(K_{q}\), one-step temporal dependence is weak.

Thus, movement through the \({K_{t} - K}_{q}\) phase space represents a change in the balance between \emph{state diversity} and \emph{temporal organization}. Movement toward higher \(K_{q}\) corresponds to increasing diversity of the states explored, whereas movement toward higher \(K_{t}\ \)corresponds to increasing uncertainty in one-step transitions. Increasing separation between \(K_{q}\) and \(K_{t}\) reflects stronger temporal dependence and greater one-step predictability, while movement toward the diagonal (\({K_{q} = K}_{t}\)) reflects progressively weaker temporal dependence. The position of a system in the \({K_{t} - K}_{q}\) phase space therefore characterizes not only the diversity of its states but also how that diversity is organized temporally.

\textbf{2.4 Terminology and levels of representation}

The \({K_{t} - K}_{q}\ \)framework can be described at three complementary levels, corresponding to its conceptual, geometric and dynamical interpretations (Table 1). To maintain terminological consistency throughout the paper, these levels are distinguished as follows.

\textbf{Table 1.} Terminology used to distinguish the conceptual, geometric and dynamical levels of the \({K_{t} - K}_{q}\) framework.

\begin{longtable}[]{@{}
  >{\raggedright\arraybackslash}p{(\columnwidth - 4\tabcolsep) * \real{0.3328}}
  >{\raggedright\arraybackslash}p{(\columnwidth - 4\tabcolsep) * \real{0.3338}}
  >{\raggedright\arraybackslash}p{(\columnwidth - 4\tabcolsep) * \real{0.3334}}@{}}
\toprule\noalign{}
\begin{minipage}[b]{\linewidth}\raggedright
Term
\end{minipage} & \begin{minipage}[b]{\linewidth}\raggedright
Meaning
\end{minipage} & \begin{minipage}[b]{\linewidth}\raggedright
Use
\end{minipage} \\
\midrule\noalign{}
\endhead
\bottomrule\noalign{}
\endlastfoot
State--transition information space & Conceptual framework combining information on ordinal-state diversity and transition uncertainty & Title and broad conceptual statements \\
(\({K_{t} - K}_{q}\)) plane & Two-dimensional geometric representation defined by (\(K_{t}\)) and (\(K_{q}\)) & Coordinates, distributions, regions, positions, and trajectories \\
(\({K_{t} - K}_{q}\)) phase space & Dynamical interpretation of the (\({K_{t} - K}_{q}\)) plane & Dynamical regimes, temporal evolution and trajectories \\
\end{longtable}

The term \emph{state--transition information space} emphasizes the information-theoretic nature of the framework: \(K_{q}\) characterizes the diversity and distribution of ordinal states, whereas \(K_{t}\ \)characterizes uncertainty in transitions between successive states. The term \({K_{t} - K}_{q}\ \)\emph{plane} refers specifically to the two-dimensional geometric representation of these quantities. The term \({K_{t} - K}_{q}\) \emph{phase space} is used when the same representation is considered as a space in which different dynamical regimes can be compared and temporal changes in dynamical organization can be followed. This use of \emph{phase space} is therefore representational rather than equivalent to the conventional phase space of dynamical-systems theory, in which coordinates are typically physical or model state variables.

\textbf{3. Canonical Systems and Phase-Space Organization}

This section evaluates the \(K_{t} - K_{q}\) information space using a set of reference time series spanning distinct dynamical regimes. The analysis demonstrates how differences in state diversity and transition uncertainty are reflected in the geometry of the phase space before the framework is applied to empirical data. The section first specifies the estimation procedure and common analysis settings, followed by an examination of the resulting phase-space locations. The finite-sample constraint is then assessed to distinguish sampling limitations from dynamically meaningful locations. Finally, where sliding-window estimates are considered, their trajectories are used to characterize temporal evolution within the phase space.

\textbf{3.1. Canonical systems and estimation procedure}

The 11 canonical time series listed in Table 2 were used as a common reference set for evaluating the \(K_{t} - K_{q}\) information space. The set comprises ordered, periodic, quasi-periodic, deterministic chaotic, correlated stochastic, and uncorrelated stochastic processes, including deterministic periodic and quasi-periodic signals, nonlinear maps, continuous-time chaotic systems, fractional Brownian motion, autoregressive processes, pink noise, and white noise. These systems were selected to provide reference points spanning distinct dynamical regimes. Their defining equations, parameter values, and abbreviations are summarized in Table 2, with the original or standard references for the individual systems cited in the text. The logistic {[}17{]} and tent maps {[}18{]} are used as canonical discrete-time nonlinear systems, while the Rössler and Lorenz systems {[}19,20{]} represent continuous-time chaotic dynamics.

Fractional Brownian motion {[}21{]} and AR(1) processes {[}22{]} provide correlated stochastic references, whereas pink noise {[}23{]} and white noise provide stochastic reference states characterized by long-range and negligible temporal correlation, respectively. The canonical white-noise realization was generated as Gaussian innovations, \({x}_{t} \sim \mathcal{N}(0,1),\ t = 1,\ldots,N\), with the \({x}_{t}\ \)independent across time, providing a standard white-noise benchmark without requiring a separate reference for its basic definition.

\textbf{Common settings.} \(N = 10,000\); \(d = 4\); \(\tau = 1\); \(A = d! = 24\). For sliding-window analyses, \(W = 1000\). For chaotic maps and continuous-time attractors, the first 5,000 observations were discarded as transients.

With these settings, the finite-sample criterion

\(K_{t}K_{q} > \frac{A - 1}{2W}\) (3.1)

gives a threshold of (0.0115). The criterion is satisfied by the canonical systems from domains II--IV and by the quasi-periodic system, whereas the constant and sinusoidal signals fall below the threshold. These two low-complexity cases lie close to the deterministic limit, where the entropy estimates are intrinsically small and more sensitive to finite-sample effects. They are therefore retained as canonical reference points, but their estimated coordinates are interpreted as limiting low-complexity locations rather than as statistically robust interior points of the phase space.

The resulting \(K_{t} - K_{q}\) coordinates are summarized in Table 2 and provide the canonical reference structure used in the phase-space organization presented in Section 3.2.

\textbf{Table 2.} Canonical time series used as reference systems for the \(K_{t} - K_{q}\) phase space.

\begin{longtable}[]{@{}
  >{\raggedright\arraybackslash}p{(\columnwidth - 8\tabcolsep) * \real{0.1727}}
  >{\raggedright\arraybackslash}p{(\columnwidth - 8\tabcolsep) * \real{0.1587}}
  >{\raggedright\arraybackslash}p{(\columnwidth - 8\tabcolsep) * \real{0.2854}}
  >{\raggedright\arraybackslash}p{(\columnwidth - 8\tabcolsep) * \real{0.2328}}
  >{\raggedright\arraybackslash}p{(\columnwidth - 8\tabcolsep) * \real{0.1504}}@{}}
\toprule\noalign{}
\begin{minipage}[b]{\linewidth}\raggedright
Canonical system
\end{minipage} & \begin{minipage}[b]{\linewidth}\raggedright
Abbreviation
\end{minipage} & \begin{minipage}[b]{\linewidth}\raggedright
Definition / generating model
\end{minipage} & \begin{minipage}[b]{\linewidth}\raggedright
Parameters
\end{minipage} & \begin{minipage}[b]{\linewidth}\raggedright
Dynamical class
\end{minipage} \\
\midrule\noalign{}
\endhead
\bottomrule\noalign{}
\endlastfoot
Constant signal & CS & \(x_{n} = C\) & \(C = 1\) & Ordered \\
Sine wave & SW & \(x_{n} = sin(\frac{2\pi f}{N})\) & \(f = 20\ cycles\) per record & Periodic \\
Quasi-periodic signal & QP & \(x_{n} = \sin\left( \frac{2\pi f_{1}}{N} \right) +\) \(\sin\left( \frac{2\pi f_{2}}{N} \right)\) & \(f_{1} = 20;f_{2} = 20\sqrt{2}\) & Quasi-periodic \\
Rössler attractor & RA & \(\dot{x} = - y - z;\dot{y} = x + ay;\)

\(\dot{z} = b + z(x - c)\) & \(a = 0.2;b = 0.2;\)

\(c = 5.7\) & Chaotic \\
Lorenz attractor & LA & \(\dot{x} = \sigma(y - x);\)

\(\dot{y} = x(\rho - z) - y;\) \(\dot{z} = xy - \beta z\) & \(\sigma = 10;\rho = 28;\)

\(\beta = 8/3\) & Chaotic \\
Logistic map & LM & \(x_{n + 1} = rx_{n}\left( 1 - x_{n} \right)\) & \(r = 3.99;x_{o} = 0.2\) & Chaotic \\
Tent map & TM & \(x_{n + 1} = 1.99x_{n},x_{n} < 1/2\)

\(1.99\left( 1 - x_{n} \right),x_{n} \geq 1/2\) & \(x_{o} = 0.37\) & Chaotic \\
Fractional Brownian motion & fBm & \(B\_ H(t)\) & \(H = 0.5\) & Stochastic \\
AR(1) process & AR1 & \(x_{n} = \varphi x_{n - 1} + \varepsilon_{n}\) & \(\varphi = 0.9;\varepsilon_{n}\mathcal{\sim N}(0,1)\) & Correlated stochastic \\
Pink noise & PN & \(S(f) \propto f^{- \alpha}\) & \(|\alpha = 1|\) & Long-range correlated stochastic \\
White noise & WN & \(x_{n} = \varepsilon_{n};\varepsilon_{n}\sim N\left( 0,\sigma^{2} \right)\) & \(\sigma^{2} = 1\) & Uncorrelated stochastic \\
\end{longtable}

\textbf{Note.} The table provides the definitions, parameter values, and dynamical classes used for the canonical reference set. Literature references for individual systems are cited in the main text.

For each canonical series, the scalar time series was transformed into a sequence of ordinal states using the Bandt--Pompe symbolic representation {[}3{]}. For an embedding dimension \(d\), the number of possible ordinal states is \(A = d!.\) Successive ordinal states were used to estimate the empirical transition matrix \(P\), whose elements are

\(P_{ij} = \ P(S_{t + 1} = \ S_{j}\ |\ S_{t} = \ S_{i}\)),

together with the ordinal-state probabilities

\[\pi_{i} = \ P\left( S_{t} = S_{i} \right).\]

The two coordinates of the information space were then calculated from the resulting state distribution and transition matrix. The horizontal coordinate, \(K_{t}\), was defined as the normalized conditional entropy of the next ordinal state given the current state,

\[K_{t} = \frac{H\left( S_{t + 1}\mid S_{t} \right)}{\log A},\]

whereas the vertical coordinate, (\(K_{q}\)), was defined as the normalized Shannon entropy of the ordinal-state distribution,

\[K_{q} = \frac{H\left( S_{t} \right)}{\log A}.\]

Both quantities are normalized to the interval {[}0,1{]}.

The same estimation settings were applied across the canonical series so that differences in the resulting (\(K_{t}\), \(K_{q}\)) coordinates reflected differences in their temporal organization rather than differences in the analysis parameters. Detailed considerations concerning finite-window estimation are provided in Supplementary Section S1.7.

\textbf{3.2. Phase-space organization of canonical systems}

The canonical time series were used to establish the empirical organization of the \(K_{t} - K_{q}\ \)phase space and to provide reference locations for the subsequent analysis of empirical data. All canonical series were evaluated using the same ordinal-symbolic settings, with embedding dimension \(d = 4\), delay \(\tau = 1\), \(A = d! = 24\ \ \)possible ordinal states, and \(N\ \)= 10,000 observations. The resulting \(K_{t}\) and \(K_{q}\) coordinates are summarized in Table 3.

\textbf{Table 3.} Canonical time series and their locations in the \(K_{t} - K_{q}\ \)phase space. Normalized temporal complexity \(K_{t}\) and structural complexity \(K_{q}\) are shown for the canonical systems used to define the four empirical domains.

\begin{longtable}[]{@{}
  >{\raggedright\arraybackslash}p{(\columnwidth - 6\tabcolsep) * \real{0.34}}
  >{\raggedright\arraybackslash}p{(\columnwidth - 6\tabcolsep) * \real{0.36}}
  >{\raggedright\arraybackslash}p{(\columnwidth - 6\tabcolsep) * \real{0.15}}
  >{\raggedright\arraybackslash}p{(\columnwidth - 6\tabcolsep) * \real{0.15}}@{}}
\toprule\noalign{}
\begin{minipage}[b]{\linewidth}\raggedright
Domain
\end{minipage} & \begin{minipage}[b]{\linewidth}\raggedright
Canonical system
\end{minipage} & \begin{minipage}[b]{\linewidth}\raggedright
\[K_{t}\]
\end{minipage} & \begin{minipage}[b]{\linewidth}\raggedright
\[K_{q}\]
\end{minipage} \\
\midrule\noalign{}
\endhead
\bottomrule\noalign{}
\endlastfoot
\multirow{3}{=}{I. Low-complexity ordered} & Constant signal (CS) & 0 & 0 \\
& Sine wave (SW) & 0.008 & 0.234 \\
& Quasi-periodic (QP) & 0.046 & 0.311 \\
\multirow{4}{=}{II. Structured nonlinear} & Rössler attractor (RA) & 0.083 & 0.580 \\
& Lorenz attractor (LA) & 0.099 & 0.631 \\
& Logistic map (LM) & 0.194 & 0.745 \\
& Tent map (TM) & 0.199 & 0.738 \\
\multirow{2}{=}{III. Highly disordered correlated stochastic} & Fractional Brownian motion (fBm) & 0.376 & 0.934 \\
& AR(1\textbf{)} process (AR1) & 0.386 & 0.947 \\
\multirow{2}{=}{IV. Near-maximally random} & 1/f (Pink) noise (PN) & 0.402 & 0.978 \\
& White noise (uniform) (WN) & 0.408 & 0.997 \\
\end{longtable}

The canonical systems span a broad portion of the admissible phase space. Their coordinates provide empirical reference points for delineating four domains, ranging from low-complexity ordered signals to near-maximally random dynamics. Figure 1 shows the distribution of the canonical systems and the corresponding empirical domains in the \(K_{t} - K_{q}\) phase space.

\includegraphics[width=4.76944in,height=4.76944in]{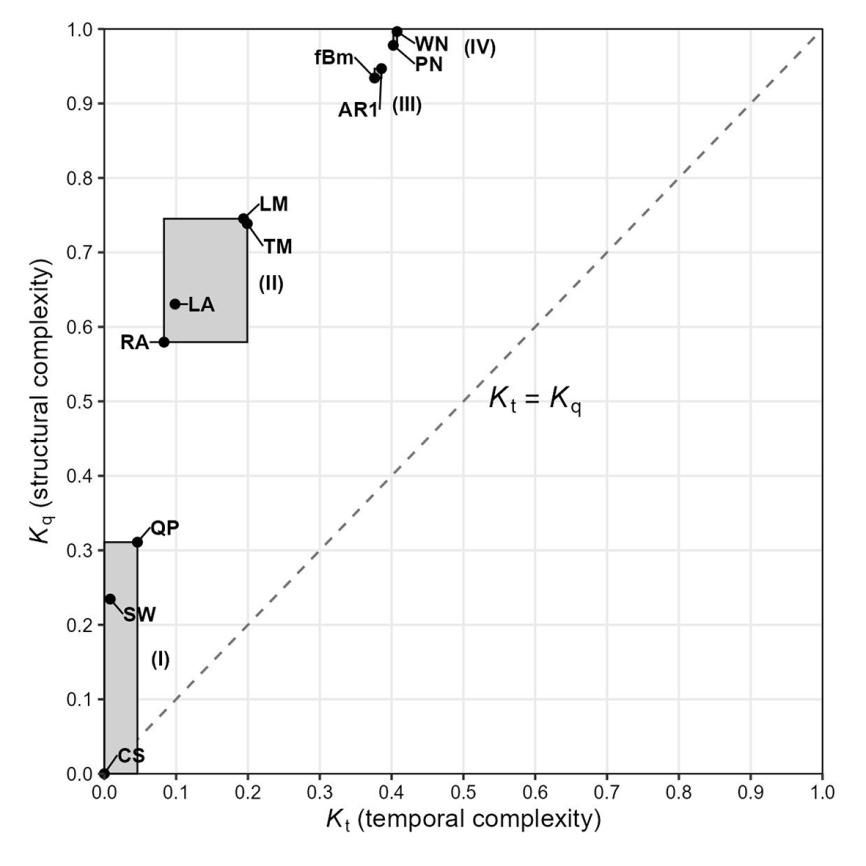}

\textbf{Figure 1.} Phase-space organization of canonical dynamical regimes\textbf{.} The \(K_{t} - K_{q}\) plane shows the locations of the canonical time series in terms of normalized temporal complexity \(K_{t}\) and structural complexity \(K_{q}\). The shaded regions denote the four empirically identified domains: (I) low-complexity ordered dynamics, (II) structured nonlinear dynamics, (III) highly disordered correlated stochastic dynamics, and (IV) near-maximally random dynamics. The diagonal \(K_{t} = K_{q}\) represents the upper admissible boundary implied by the entropy inequality \(H\left( S_{t + 1}\mid S_{t} \right) \leq \ H\left( S_{t} \right)\). Points indicate the \(K_{t} - K_{q}\) coordinates obtained from the canonical series, while the regional boundaries are determined empirically from their observed distributions.

Four empirical domains are distinguished from the distribution of the canonical coordinates. \emph{Domain I (low-complexity ordered)} is characterized by 0\(\  \leq K_{t}\) \(\leq\) 0.046 and 0 \(\leq K_{q}\) \(\leq\) 0.311. It contains the constant, sinusoidal, and quasi-periodic series, which occupy the lower-left portion of the phase space, with both temporal uncertainty and state diversity remaining relatively low.

\emph{Domain II (structured nonlinear)} is characterized by 0.083 \(\leq K_{t}\) \(\leq\) 0.199 and 0.580 \(\leq K_{q}\) \(\leq\) 0.745. The tent map, logistic map, Lorenz system, and Rössler system are located in this domain, with substantially larger values of both \(K_{t}\ \)and \(K_{q}\) than the ordered signals.

\emph{Domain III (highly disordered correlated stochastic)} is characterized by 0.376 \(\leq K_{t}\) \(\leq\) 0.386 and 0.934\(\leq K_{q}\) \(\leq\) 0.947. The fractional Brownian motion and AR(1) series occupy this domain, exhibiting high values of both coordinates.

\emph{Domain IV (near-maximally random)} is characterized by 0.402\(\leq K_{t}\) \(\leq\) 0.408 and 0.978 \(\leq K_{q}\) \(\leq\) 0.997, corresponding to the canonical systems located closest to the maximally disordered corner of the phase space.

The individual canonical coordinates further illustrate the separation among these regions. The constant series has (\(K_{t},K_{q}\))=(0,0), while the sinusoidal and quasi-periodic series have (0.008, 0.234) and (0.046, 0.311), respectively. The tent map and logistic map are located at approximately (0.199, 0.738) and (0.194, 0.745), respectively. The Rössler and Lorenz systems are located at approximately (0.083, 0.580) and (0.099, 0.631), respectively, whereas the fractional Brownian motion and AR(1) series occupy approximately (0.376, 0.934) and (0.386, 0.947). Pink-noise and white series occupy approximately (0.402, 0.978) and (0.408, 0.997), respectively.

Thus, the canonical series define four empirically distinguishable domains within the admissible \(K_{t} - K_{q}\ \)phase space, ranging from low-complexity ordered signals to near-maximally random dynamics. These domains are not arbitrary subdivisions of the \(K_{t} - K_{q}\) plane; rather, they are identified from the locations and distribution of the canonical systems listed in Table 3 and provide the reference structure for the subsequent analysis. Accordingly, the four domains should be interpreted as an \emph{empirical organization} of the phase space rather than as formal classes or universal theoretical classification boundaries.

\textbf{3.3. Phase-space trajectories of canonical systems}

To examine the temporal character of the \(K_{t} - K_{q}\) plane representation, sliding-window estimates were applied to representative canonical systems from each of the four empirical domains identified in Section 3.2. For each series, the total record length was \(N\) = 10,000, and a window of length \(W\) = 1000 was advanced in increments of 25 observations. Each window therefore provides a local estimate of the normalized temporal complexity \(K_{t}\) and structural complexity \(K_{q}\), yielding a time-ordered trajectory through the (\(K_{t},K_{q}\)) phase space as the window moves along the time series. The trajectories and their temporal progression are shown in Fig. 2.

\includegraphics[width=6.5in,height=5.30347in]{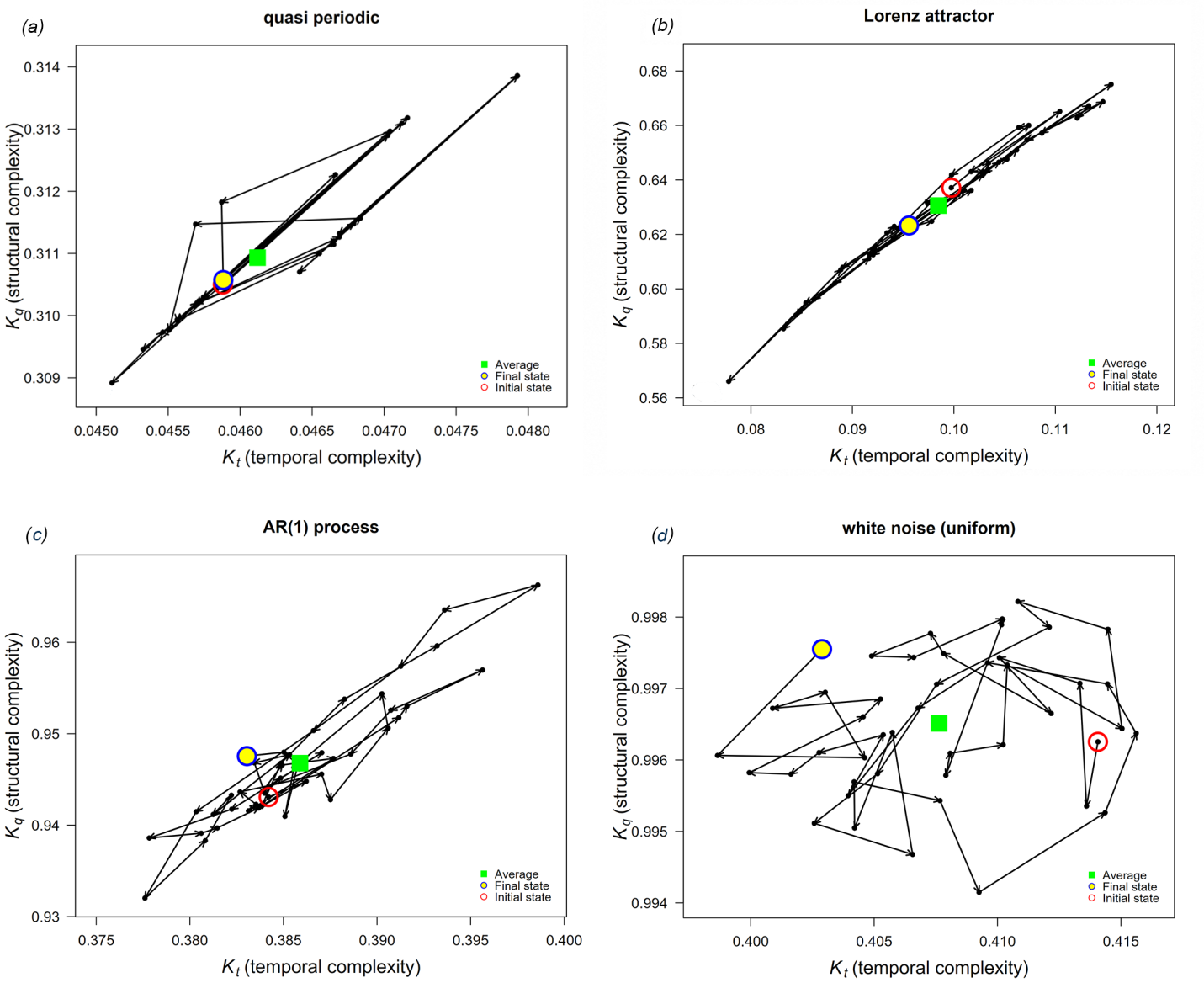}

\textbf{Figure 2.} Trajectories of canonical dynamical systems in the \(K_{t} - K_{q}\ \)plane, where \(K_{t}\) represents temporal complexity and \(K_{q}\) structural complexity: (a) periodic system, (b) Lorenz attractor, (c) AR(1) process, and (d) uniform white noise. Black arrows indicate the temporal evolution of the state through successive points. The red open circle denotes the initial state, the yellow circle the final state, and the green square the mean position of the trajectory. The trajectories illustrate distinct patterns of dynamical organization and variability in the \(K_{t} - K_{q}\)phase space.

For the analysis of temporal trajectories, one representative canonical system was selected from each empirical domain. The quasi-periodic (QP) series was chosen for Domain I (low-complexity ordered dynamics), the Lorenz attractor (LA) for Domain II (structured nonlinear dynamics), the AR(1) process for Domain III (highly disordered correlated stochastic dynamics), and white noise (WN) for Domain IV (near-maximally random dynamics). These four systems were selected because each represents a distinct empirically identified dynamical domain, while together they span the progression from ordered to near-maximally random dynamics. This selection provides a direct comparison of \(K_{t} - K_{q}\ \)trajectories across qualitatively distinct dynamical regimes without introducing unnecessary redundancy.

The resulting \(K_{t} - K_{q}\) trajectories illustrate how the temporal and structural characteristics of the systems evolve within the phase space. The trajectories occupy distinct regions corresponding to their respective dynamical regimes, providing a continuous representation of the transition from low-complexity ordered behavior through structured nonlinear and disordered stochastic dynamics to near-maximal randomness.

The quasi-periodic system (Fig. 2a) occupies a low-\(K_{t}\), intermediate-\({\ K}_{t\ }\)position in the\(\ K_{t} - K_{q}\) phase space, with approximately (\({\ K}_{t\ }\)= 0.046) and (\({\ K}_{q\ }\)= 0.311). The low \({\ K}_{t\ }\) indicates that transitions between successive ordinal states remain strongly constrained and therefore relatively predictable, whereas the intermediate \({\ K}_{q\ }\)reflects a broader repertoire of ordinal states and correspondingly greater structural complexity. The resulting trajectory therefore remains concentrated in the low-\(K_{t}\), intermediate-\({\ K}_{q\ }\)region of the phase space, with limited temporal transition complexity and moderate structural complexity.

Figure 2b shows the evolution of the Lorenz system in the \(K_{t} - K_{q}\) phase space. The trajectory occupies an elongated region extending approximately from\(\ (K_{t}\)= 0.078, \({\ K}_{q}\) = 0.566) to (\({\ K}_{t}\) =0.115,\({\ K}_{q}\)= 0.675). Most states are concentrated along an upward trend, indicating that temporal complexity (\({\ K}_{t}\)) and structural complexity (\({\ K}_{q}\)) tend to increase together. Nevertheless, the trajectory exhibits several reversals and local excursions, demonstrating substantial variation in the complexity coordinates during the evolution. The initial state is located near \({\ (K}_{t} \approx\)0.078), and (\({\ K}_{q} \approx\) 0.566), while the average state is at approximately (\({\ K}_{t} \approx\)0.098), and (\({\ K}_{q} \approx\) 0.631).

Figure 2c shows the evolution of the AR(1) process in the \(K_{t} - K_{q}\) phase space. The trajectory occupies a relatively compact but extended region, with \(K_{t}\ \)values ranging approximately from 0.378 to 0.399 and \(K_{q}\) values from 0.932 to 0.966. Most states are concentrated around (\(K_{t}\) \(\approx\) 0.384-0.391) and (\(K_{q}\) \(\approx\) 0.94-0.95), while several excursions extend toward higher values of both complexity measures. The trajectory exhibits an overall upward tendency, indicating that increases in temporal complexity are generally accompanied by increases in structural complexity. However, several reversals and local excursions are evident, demonstrating temporal variability in the \({(K}_{t},K_{q})\) coordinates. The initial state is located near (\(K_{t}\) \(\approx\) 0.384) and (\(K_{q} \approx\) 0.943), while the average state is approximately (\(K_{t}\) \(\approx\) 0.386) and (\(K_{q}\) \(\approx\) 0.947).

Figure 2d shows the evolution of white noise in the \(K_{t} - K_{q}\) phase space. The trajectory occupies a high-complexity region, with \(K_{t}\)values of approximately 0.399--0.416 and \(K_{q}\) values of 0.994--0.998. The states are distributed close to the upper part of the phase space, with pronounced local reversals and excursions. The average state is approximately (\(K_{t}\)= 0.408), (\(K_{q}\)= 0.996).

\textbf{4. Empirical application to U.S. river flow records}

The empirical analysis used monthly naturalized streamflow records from 1,879 U.S. rivers covering the period 1950--2015. The data were obtained from the U.S. Geological Survey (USGS) ScienceBase Catalog and are based on simulated streamflow for 2,622,273 stream reaches defined by the National Hydrography Dataset (NHD) Version 2.0 across the continental United States. The streamflow simulations were generated using a random forest ensemble approach {[}24{]}. For the present analysis, we selected gauge stations located at the outlets of individual watersheds, where the mainstem represents the integrated flow from upstream tributaries. Accordingly, the analysis was based on naturalized monthly streamflow from 1,879 river outlets. For each river, ordinal patterns with embedding dimension \(d = 4\ \)and time delay \(\tau = 1\) were used to construct the transition matrix. The normalized transition entropy rate \(K_{t}\) and normalized Shannon entropy \(K_{q}\) were then calculated from the resulting transition probabilities and ordinal-state probabilities, respectively, providing one (\(K_{t},K_{q}\ \)) coordinate for each river record.

The 1,879 gauge stations were grouped into five drainage-area classes, ranging from very small ((\(\leq\)100) km\(^2\)) to very large ((\textgreater100,000) km²). The distribution of gauge stations among these classes is summarized in Table 4. The medium drainage-area class contains the largest number of stations (1,144; 60.9\%), followed by the large (408; 21.7\%) and small (203; 10.8\%) classes, whereas the very small and very large classes comprise 49 (2.6\%) and 75 (4.0\%) stations, respectively.

\textbf{Table 4.} Distribution of gauge stations used in the study by drainage-area size.

\begin{longtable}[]{@{}
  >{\raggedright\arraybackslash}p{(\columnwidth - 6\tabcolsep) * \real{0.2383}}
  >{\raggedright\arraybackslash}p{(\columnwidth - 6\tabcolsep) * \real{0.3656}}
  >{\raggedright\arraybackslash}p{(\columnwidth - 6\tabcolsep) * \real{0.2953}}
  >{\raggedright\arraybackslash}p{(\columnwidth - 6\tabcolsep) * \real{0.1008}}@{}}
\toprule\noalign{}
\begin{minipage}[b]{\linewidth}\raggedright
Category
\end{minipage} & \begin{minipage}[b]{\linewidth}\raggedright
Drainage area
\end{minipage} & \begin{minipage}[b]{\linewidth}\raggedright
Number of gauge stations
\end{minipage} & \begin{minipage}[b]{\linewidth}\raggedright
\%
\end{minipage} \\
\midrule\noalign{}
\endhead
\bottomrule\noalign{}
\endlastfoot
Very small & (\(\leq\)100) km\(^2\) & 49 & 2.6 \\
Small & 100--1000 km² & 203 & 10.8 \\
Medium & 1000--10000 km² & 1144 & 60.9 \\
Large & 10000--100000 km² & 408 & 21.7 \\
Very large & (\textgreater100000 km²) & 75 & 4.0 \\
\end{longtable}

\includegraphics[width=6.42928in,height=3.40625in]{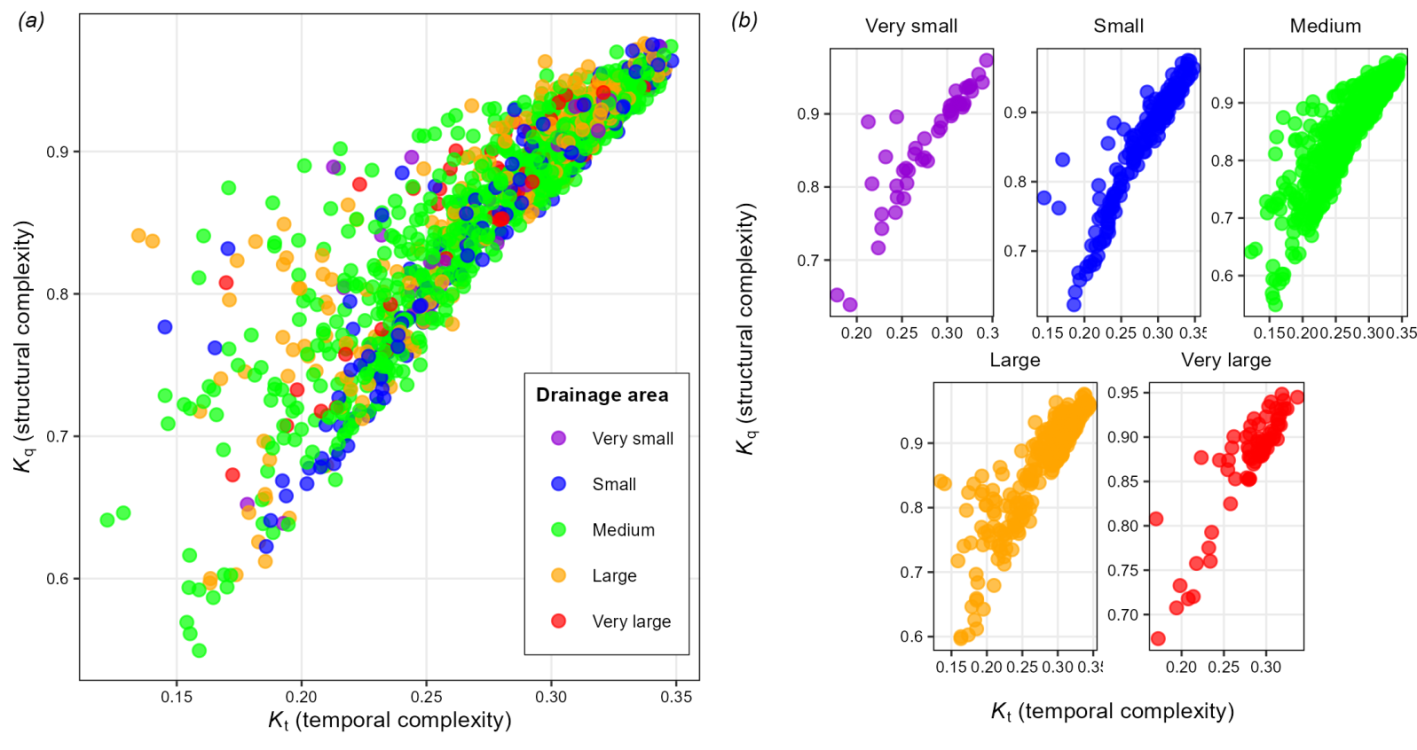}

\textbf{Figure 3.} Distribution of U.S. river records in the \(K_{t} - K_{q\ }\)plane according to drainage-area size. \textbf{(a)} All 1,879 river records shown together and distinguished by five drainage-area classes: very small, small, medium, large, and very large. \textbf{(b)} The same records shown separately for each drainage-area class, illustrating the within-class distributions in the \(K_{t} - K_{q}\ \)plane.

The corresponding distribution of the river records in the \(K_{t} - K_{q\ }\) plane is shown in Figure 3. Figure 3a presents all 1,879 records simultaneously, distinguished by drainage-area class, whereas Figure 3b shows the distributions separately for each class. The figure therefore provides an empirical characterization of the organization of river records in the \(K_{t} - K_{q\ }\)plane and its dependence on drainage-area size.

The complete set of river records occupies a well-defined intermediate region of the \(K_{t} - K_{q\ }\)plane, approximately spanning (0.03 \textless{}\(\ K_{t}\ \)\textless{} 0.14) and (0.40 \textless{}\({\ K}_{q\ }\)\textless{} 0.70) (Fig. 3a). The river records are therefore distinct from the near-ordered and near-random regimes represented by the canonical systems considered above. The distribution is elongated from the lower-left towards the upper-right, indicating a positive association between \(K_{t}\) and \(K_{q\ }\), such that records with larger transition entropy rates generally also exhibit larger Shannon entropy.

The class-specific distributions (Fig. 3b) show systematic differences in both position and spread. Very small rivers occupy a relatively restricted region of the lower-\(K_{t}\), lower-\(K_{q\ }\) portion of the plane. Small and medium rivers extend over progressively broader ranges of both quantities, with the medium drainage-area class showing the greatest overall spread. Large and very large rivers are more concentrated towards the upper-\(K_{t}\), upper-\(K_{q\ }\) region. Although the five classes overlap substantially, their distributions show a progressive displacement from the lower-left towards the upper-right with increasing drainage-area size.

This organization indicates that drainage-area size is systematically associated with the dynamical characteristics represented by \(K_{t}\) and \(K_{q\ }\). Increasing drainage area is associated with a shift towards larger values of both quantities, rather than with a change confined to a single coordinate. The observed pattern therefore indicates scale-dependent organization of river dynamics in the \(K_{t} - K_{q}\) plane. Because the analysis is based on monthly streamflow, this result characterizes the organization of river dynamics at the monthly temporal scale and does not, by itself, imply a universal scaling law across spatial or temporal scales.

Here, the \(K_{t} - K_{q\ }\ \)plane denotes the static distribution of the empirical river records, with each record represented by a single (\(K_{t},K_{q}\)) coordinate. The term \(K_{t} - K_{q\ }\) phase space is reserved for the broader dynamical framework in which successive states can be represented as trajectories through this plane.

\textbf{5. Discussion}

\emph{The} \(K_{t} - K_{q}\) \emph{phase space as a representation of dynamical organization}. The \(K_{t} - K_{q}\ \)phase space provides a means of distinguishing the diversity of ordinal states from temporal unpredictability, rather than representing dynamical complexity by a single scalar quantity. \(K_{q}\ \)quantifies the diversity and distribution of ordinal states, whereas \(K_{t}\ \)quantifies the uncertainty associated with transitions between successive ordinal states. Their joint representation therefore distinguishes systems that may exhibit a similarly rich collection of ordinal states but substantially different temporal organization. Because the transition entropy is bounded by the corresponding state entropy, \(K_{t} \leq K_{q}\), the relative positions of \(K_{t}\) and \(K_{q}\ \)and particularly their separation, provide information about dynamical organization. A system may explore a relatively rich repertoire of ordinal states, while retaining strong temporal organization, resulting in a relatively low \(K_{t}\) but appreciable \(K_{q}\). Conversely, systems with weaker temporal organization tend to exhibit larger \(K_{t}\ \)values together with high \(K_{q}\) Thus, the \(K_{t} - K_{q}\) representation captures complementary aspects of dynamical behavior: the diversity of accessible ordinal states and the uncertainty of their temporal transitions.

This distinction is particularly evident in quasi-periodic dynamics. A quasi-periodic signal contains several incommensurate frequencies and therefore does not repeat exactly, while remaining deterministic and strongly organized (Fig. 2a). Its position in the low-\(K_{t}\), intermediate-\(K_{q}\ \)region reflects a richer collection of ordinal states than that of a strictly periodic signal, without a comparable increase in transition uncertainty (Fig. 1). In this sense, quasi-periodicity increases the diversity of accessible states while preserving a relatively high degree of temporal organization. It therefore provides a clear example of how the \(K_{t} - K_{q}\) representation can distinguish increased structural richness from a corresponding increase in temporal unpredictability.

\emph{Dynamical domains and temporal trajectories}. The canonical systems demonstrate that the \(K_{t} - K_{q}\) phase space is not merely a static characterization of dynamical regimes but can also describe their temporal evolution. The four empirical domains provide reference regimes extending from low-complexity ordered dynamics through structured nonlinear and highly disordered, correlated stochastic dynamics to near-maximally random behavior. Their separation establishes a common framework within which different forms of temporal organization can be compared.

The sliding-window trajectories further show that a dynamical system does not necessarily remain at a single point in this plane. For the Lorenz attractor (Fig. 2b), the trajectory exhibits an overall upward orientation, indicating that ordinal-state diversity and transition uncertainty tend to vary together, while local reversals demonstrate that this relationship is not strictly monotonic. The trajectory therefore captures temporal changes in the dynamical organization of the attractor that would be lost if only its mean \(K_{t} - K_{q}\) position were considered. The mean position characterizes the dominant dynamical regime, whereas the trajectory reveals temporal fluctuations around that regime.

A qualitatively different behavior is observed for the AR(1) process (Fig. 2c). Its trajectory remains within the high-\(K_{t}\), high-\(K_{q\ }\)region but fluctuates substantially within this domain. The high \(K_{q\ }\) values indicate a broad repertoire of ordinal patterns, while the relatively high \(K_{t}\) values indicate substantial uncertainty in transitions between successive states. The trajectory therefore characterizes a highly disordered, correlated stochastic regime, while retaining temporal variability associated with its correlation structure.

White noise (Fig. 2d) provides the limiting case. Its trajectory is concentrated close to the upper limits of both \(K_{t}\ \)and \(K_{q\ }\), consistent with near-maximal ordinal-state diversity and transition uncertainty. The fluctuations and reversals within this region reflect the absence of persistent temporal ordering, while the proximity of \(K_{q}\ \)to unity indicates an approximately uniform distribution of ordinal patterns. Together, these canonical trajectories demonstrate that the \(K_{t} - K_{q}\) framework can distinguish both dynamical regimes and temporal variability within those regimes.

These examples establish an important distinction between \emph{where a system is located} in the \(K_{t} - K_{q}\ \)phase space and \emph{how it moves through that space.} The former characterizes its dominant dynamical regime, whereas the latter describes temporal changes in its dynamical organization.

\emph{River dynamics occupy an intermediate dynamical regime}. The 1,879 U.S. river records occupy a distinct intermediate region of the \(K_{t} - K_{q}\ \)plane, separated from both the low-complexity ordered systems and the highly disordered stochastic systems represented by the canonical domains. Their intermediate \(K_{t}\ \)and \(K_{q}\ \)values indicate that river discharge dynamics exhibit neither the low transition uncertainty characteristic of strongly ordered systems nor the near-maximal transition uncertainty associated with white noise. The river population therefore occupies an intermediate regime in terms of the statistical organization captured by the two entropy measures.

An important complementary result is provided by the Lyapunov-exponent analysis of the same 1,879 U.S. river records reported by {[}25{]}. All river time series exhibited positive Lyapunov exponents, with the exponents calculated using the methodology described in that study. This provides independent evidence of sensitivity to initial conditions across the entire river ensemble. The positive Lyapunov exponents therefore strengthen the interpretation of the intermediate position of the rivers in the \(K_{t} - K_{q}\ \)plane: their dynamics exhibit measurable dynamical instability, while remaining distinct from the near-maximally disordered regime represented by white noise.

The agreement between the Lyapunov-exponent results and the \(K_{t} - K_{q}\ \)representation is informative, because the two approaches characterize different aspects of the dynamics. The Lyapunov exponent describes the rate of divergence of nearby trajectories and hence sensitivity to initial conditions, whereas the \(K_{t} - K_{q}\ \)coordinates describe the diversity of ordinal states and the uncertainty of their temporal transitions. The two measures should therefore not be regarded as equivalent. Rather, their combination indicates that river discharge dynamics can exhibit sensitivity to initial conditions, while retaining a structured statistical organization that is neither strongly ordered nor maximally random.

The intermediate location of the rivers is consistent with discharge dynamics resulting from the interaction of temporally organized processes and multiple sources of variability. Potential contributors include precipitation forcing, storage and release within catchments, catchment heterogeneity, groundwater interactions, and the integration of spatially distributed inputs across drainage basins. The \(K_{t} - K_{q}\ \)distribution should not, however, be interpreted as evidence for a single underlying mechanism. Rather, it characterizes the resulting dynamical organization of the aggregated discharge records and allows their empirical states to be compared with the canonical reference regimes without assigning them to a specific deterministic or stochastic model.

\emph{Drainage-area dependence and scale-dependent organization}. A systematic organization is also evident when the river records are considered according to drainage-area size. Smaller drainage areas tend to occupy the lower-\(K_{q}\), lower-\(K_{q}\) portion of the river distribution, whereas progressively larger drainage areas tend to shift toward higher values of both quantities. Thus, river position in the \(K_{t} - K_{q}\ \)plane changes systematically with drainage-area size, indicating that the dynamical organization of streamflow is scale-dependent.

This pattern is consistent with the fact that increasing drainage area changes the spatial scale over which hydrological processes are integrated. Larger basins involve stronger spatial averaging, which can suppress some local fluctuations. At the same time, however, they incorporate a greater diversity of climatic, geomorphological, land-surface and hydrological conditions. The resulting increase in spatial heterogeneity and in the number of interacting processes can counteract the effect of averaging and contribute to the observed displacement toward higher \(K_{t}\) and \(K_{q}\).

The observed organization should therefore not be interpreted as a simple increase in complexity caused by basin size alone. Rather, drainage area changes the degree of spatial integration and the number and diversity of interacting processes represented in the discharge record. The \(K_{t} - K_{q}\) plane makes this scale dependence visible as an organized displacement of river populations across dynamical space.

This result is important because it indicates that the dynamical characteristics of river discharge cannot necessarily be regarded as scale-invariant properties of a basin. Instead, they depend on the spatial scale at which the hydrological system is observed and aggregated. In this sense, the \(K_{t} - K_{q}\ \)framework provides a way of characterizing \emph{scale-dependent dynamical organization}, rather than implying a conventional scaling law or power-law relationship.

\emph{Temporal resolution, predictability and methodological dependence.} The scale dependence observed among rivers also has a temporal counterpart. The present analysis uses monthly discharge observations and therefore characterizes river dynamics at a substantially longer temporal resolution than analyses based on daily or sub-daily observations. Processes that dominate at shorter time scales may be averaged out at the monthly scale, whereas slower storage, persistence and seasonal processes become more prominent. Consequently, the position of a river in the \(K_{t} - K_{q}\ \)plane should be understood as a property of the river dynamics at the temporal resolution and sampling scheme used in the analysis, rather than as an invariant characteristic of the river itself.

The use of sliding windows introduces a further scale into the analysis. Window length determines the temporal extent over which the local ordinal-state distribution and transition structure are estimated. Shorter windows can reveal more rapid changes in dynamical organization but are more strongly affected by finite-sample variability, whereas longer windows provide more stable estimates at the cost of temporal resolution. The observed trajectories should therefore be interpreted in relation to both the physical temporal scale of the discharge process and the statistical scale imposed by the analysis.

These considerations also have implications for predictability. \(K_{t}\ \)is directly related to uncertainty in the temporal evolution of ordinal states and therefore provides information complementary to a measure based solely on the distribution of states. A systematic displacement toward higher \(K_{t}\ \)with drainage area indicates increasing transition uncertainty within the monthly observations, although \(K_{t}\ \)should not be interpreted as a direct estimate of a forecast horizon. Predictability depends additionally on the prediction model, temporal resolution, forcing, and state information available to the forecaster. The value of the \(K_{t} - K_{q}\) framework is instead that it provides a model-independent characterization of how temporal uncertainty is organized relative to structural complexity.

The positive Lyapunov exponents reported for all 1,879 river records provide an additional perspective on this predictability issue. Because a positive Lyapunov exponent is associated with exponential divergence of nearby trajectories, the result indicates that sensitivity to initial conditions is widespread across the river ensemble. However, the Lyapunov exponent and \(K_{t}\ \)should not be regarded as equivalent measures: the former concerns the rate of divergence in reconstructed dynamical trajectories, whereas the latter characterizes the statistical uncertainty of ordinal-state transitions. Their complementary information is therefore more informative than either measure considered alone.

The quantitative boundaries of the canonical domains should likewise be regarded as empirical reference regions rather than universal thresholds. Their precise locations can depend on embedding dimension, time delay, window length, sample size, and the ordinal-state representation. Nevertheless, the persistence of distinct dynamical regimes across the canonical examples and their separation from the river population support the usefulness of the framework as a comparative representation of dynamical organization.

\emph{Limitations and broader implications}. Several limitations should be recognized. First, the present river analysis is based on monthly discharge records and therefore does not characterize the full range of temporal scales present in hydrological systems. Analyses using daily or sub-daily observations may produce different \(K_{t} - K_{q}\ \)distributions and trajectories. Second, the estimates depend on methodological choices such as embedding dimension, delay, window length and sampling density. These parameters should therefore be considered explicitly when comparing different datasets or applications.

Third, the empirical domains defined from canonical systems should not be interpreted as rigid universal boundaries between dynamical classes. They provide reference regimes within the present normalization and parameterization. Future studies involving larger ensembles of deterministic, stochastic, and observational systems could refine these boundaries and establish their statistical robustness.

Despite these limitations, the results demonstrate that the \(K_{t} - K_{q}\ \)framework can provide information that is not contained in a single scalar measure. The canonical examples establish reference dynamical regimes, while the sliding-window trajectories show how dynamical organization varies around those regimes. In the river application, the framework reveals both an intermediate dynamical regime and a systematic displacement through the \(K_{t} - K_{q}\ \)plane with increasing drainage area.

More importantly, the river results show that this statistical organization is complementary to evidence of dynamical instability. The positive Lyapunov exponents found for all 1,879 records indicate widespread sensitivity to initial conditions, whereas the \(K_{t} - K_{q}\ \)representation identifies the statistical organization associated with that instability and its variation across spatial scale. These results support a view of river discharge dynamics in which \emph{instability, temporal organization and state diversity coexist}, rather than being represented adequately by a single measure of complexity.

More generally, the \(K_{t} - K_{q}\ \)phase space provides a compact framework for comparing dynamical regimes, temporal variability and scale-dependent organization across natural systems. Its usefulness can be assessed in other applications by examining whether the joint \(K_{t}\ \)and \(K_{\ q}\)distributions, together with their temporal trajectories, reveal systematic dynamical organization across relevant spatial or temporal scales.

\textbf{6. Conclusions}

This study introduced the \(K_{t} - K_{q}\) plane as a representation of dynamical organization that distinguishes the diversity of ordinal states from the uncertainty of their temporal transitions. The canonical systems demonstrated that these two quantities provide complementary information: systems with similarly rich ordinal-state distributions can differ substantially in their temporal organization, while sliding-window trajectories reveal temporal changes that are not represented by a single mean coordinate. The resulting canonical domains provide empirical reference regimes spanning low-complexity ordered dynamics, structured nonlinear dynamics, highly disordered correlated stochastic dynamics, and near-maximally random dynamics.

Application to 1,879 U.S. river records showed that monthly streamflow dynamics occupied a distinct intermediate region of the \(K_{t} - K_{q}\) plane, separated from both the strongly ordered and highly disordered reference regimes. More importantly, the river records showed a systematic displacement toward higher \(K_{t}\)and \(K_{q}\)with increasing drainage area. This organization indicates that the dynamical characteristics of streamflow change with the spatial scale of basin integration, rather than remaining invariant across drainage areas. The result therefore indicates scale-dependent organization in the \(K_{t} - K_{q}\) plane, rather than a conventional scaling or power-law relation.

The analysis also establishes that the coordinates and trajectories should be interpreted jointly. A position in the plane identifies the prevailing dynamical organization, whereas movement through the plane describes temporal variation in that organization. This distinction is particularly relevant for observational systems whose dynamics cannot be adequately represented by a single stationary complexity value.

The \(K_{t} - K_{q}\) framework consequently provides a quantitative basis for comparing dynamical organization across systems and observational scales. Its principal value is not the assignment of systems to fixed complexity classes, but the identification of how state diversity, transition uncertainty, and their temporal variation are jointly organized. For hydrological applications, extending the analysis across daily, sub-daily, and monthly observations, together with systematically varied window lengths and basin scales, would provide a systematic means of investigating how dynamical organization changes across temporal and spatial scales.

\hypertarget{section}{%
\paragraph{}\label{section}}

\textbf{Data accessibility.} Data and code supporting the paper can be found at \url{https://doi.org/10.17605/OSF.IO/G7P2Z}.

\textbf{Declaration of AI use}. AI was used for minor editing tasks to improve readability and language. After using AI, the authors reviewed and edited the content as needed and assume full responsibility for the publication\textquotesingle s content.

\textbf{Authors\textquotesingle{} contributions.} D.T.M.~conceptualization, formal analysis, investigation, methodology, data curation, project administration, resources, software, visualization, supervision, validation, writing - original draft; writing - review and editing; V.P.S.~conceptualization, methodology, formal analysis, supervision, validation, funding acquisition, writing - review and editing; S.M-M. investigation, methodology, data curation; validation, visualization, writing - review and editing. All authors gave final approval for publication and agreed to be held accountable for the work performed therein.

\textbf{Conflict of interest declaration.} We declare that we have no competing interests.

\textbf{Funding:} This work was supported by the Ministry of Science, Techno­logical Development and Innovation of the Republic of Serbia {[}grant number 451-03-33/2026-03/200172{]}.

\textbf{References:}

1. Shannon CE. 1948 A mathematical theory of communication. \emph{Bell Syst. Tech. J.} \textbf{27}, 379--423. (doi:10.1002/j.1538-7305.1948.tb01338.x)

2. Cover TM, Thomas JA. 2006 \emph{Elements of Information Theory}, 2nd edn. Hoboken, NJ: Wiley.

3. Bandt C, Pompe B. 2002 Permutation entropy: A natural complexity measure for time series. \emph{Phys. Rev. Lett.} \textbf{88}, 174102. (doi:10.1103/PhysRevLett.88.174102)

4. Rosso OA, Larrondo HA, Martín MT, Plastino A, Fuentes MA. 2007 Distinguishing noise from chaos. \emph{Phys. Rev. Lett.} \textbf{99}, 154102. (doi:10.1103/PhysRevLett.99.154102)

5. Zunino L, Zanin M, Tabak BM, Pérez DG, Rosso OA. 2012 Complexity--entropy causality plane: A useful approach to quantify the stock market inefficiency. \emph{Physica A} \textbf{389}, 1891-1901. (doi:10.1016/j.physa.2010.01.007)

6. Keller K, Sinn M. 2005 Ordinal analysis of time series. \emph{Physica A} \textbf{356}, 114--120. (doi:10.1016/j.physa.2005.05.022)

7. Politi A, Ricci L. 2025 Improved reconstruction of chaotic signals from ordinal networks. \emph{Entropy} \textbf{27}, 499. (doi:10.3390/e27050499)

8. Salas C, Durán O, Vergara JI, Arata A. 2024 Permutation entropy: An ordinal pattern-based resilience indicator. \emph{Entropy} \textbf{26}, 961. (doi:10.3390/e26110961)

9. Roy O, Campbell-Cousins A, Fabila Carrasco JS, Parra MA, Escudero J. 2024 Graph permutation entropy: Extensions to the continuous case, a step towards ordinal deep learning, and more. \emph{arXiv preprint arXiv:2407.07524}.

10. Zhao X, Ji M, Zhang N, Shang P. 2020 Permutation transition entropy: Measuring the dynamical complexity of financial time series. \emph{Chaos Solitons Fractals} \textbf{139}, 110019. (doi:10.1016/j.chaos.2020.109962)

11. Zhang B, Shang P, Liu J. 2021 Transition-based complexity-entropy causality diagram: A novel method to characterize complex systems. \emph{Commun. Nonlinear Sci. Numer. Simul.} \textbf{95}, 105660. (doi:10.1016/j.cnsns.2020.105660)

12. Chen X, Xu G, He B, Zhang S, Su Z, Jia Y, Zhang X, Zhao Z.~ 2023 Capturing synchronization with complexity measure of ordinal pattern transition network constructed by crossplot. \emph{R. Soc. Open Sci.} \textbf{10}, 230500. (doi:10.1098/rsos.221067)

13. Almendral JA, Leyva I, Sendiña-Nadal I. 2023 Unveiling the connectivity of complex networks using ordinal transition methods. \emph{Entropy} \textbf{25}, 1079. (doi:10.3390/e25071079)

14. Chen Y, Ling G, Song X, Tu W. 2023. Characterizing the statistical complexity of nonlinear time series via ordinal pattern transition networks. Physica A: Statistical Mechanics and its Applications, \textbf{618}, 128702. (\href{https://doi.org/10.1016/j.physa.2023.128670}{doi:10.1016/j.physa.2023.128670})

15. Crutchfield JP, Young K. 1989 Inferring statistical complexity. \emph{Phys. Rev. Lett.} \textbf{63}, 105--108. (doi:10.1103/PhysRevLett.63.105)

16. Norris JR. 1997 \emph{Markov Chains}. Cambridge, UK: Cambridge University Press.

17. May RM. 1976 Simple mathematical models with very complicated dynamics. \emph{Nature} \textbf{261}, 459--467. (doi:10.1038/261459a0)

18. Alligood KT, Sauer TD, Yorke JA. 1996 \emph{Chaos: An Introduction to Dynamical Systems}. New York, NY: Springer.

19. Lorenz EN. 1963 Deterministic nonperiodic flow. \emph{J. Atmos. Sci.} \textbf{20}, 130--141. (doi:10.1175/1520-0469(1963)020\textless0130:DNF\textgreater2.0.CO;2)

20. Rössler OE. 1976 An equation for continuous chaos. \emph{Phys. Lett. A} \textbf{57}, 397--398. (doi:10.1016/0375-9601(76)90101-8)

21. Mandelbrot BB, Van Ness JW. 1968 Fractional Brownian motions, fractional noises and applications. \emph{SIAM Rev.} \textbf{10}, 422--437. (doi:10.1137/1010093)

22. Brockwell PJ, Davis RA. 1991 \emph{Time Series: Theory and Methods}, 2nd edn. New York, NY: Springer-Verlag.

23. Kasdin NJ. 1995 Discrete simulation of colored noise and stochastic processes and 1/f\(^{\alpha}\) power-law noise generation. \emph{Proc. IEEE} \textbf{83}, 802--827. (doi:10.1109/5.381848)

24. Miller MP, Carlisle DM, Wolock DM, Wieczorek M. 2018 A database of natural monthly streamflow estimates from 1950 to 2015 for the conterminous United States. \emph{J. Am. Water Resour. Assoc.} \textbf{54}, 1258--1269. (doi:10.1111/1752-1688.12685)

25. Mihailović DT, Malinović-Milićević S, Han J, Singh VP. 2023 Complexity and chaotic behavior of the U.S. Rivers and estimation of their prediction horizon. \emph{J. Hydrol.} \textbf{622}, 129730. (doi:10.1016/j.jhydrol.2023.129730)

\clearpage
\appendix
\section*{Supplementary Material S1. Mathematical and Estimation Details}
\addcontentsline{toc}{section}{Supplementary Material S1. Mathematical and Estimation Details}

\subsection*{S1.1 Entropy identities}

Let \(S_t\) denote the ordinal state associated with the embedded vector at time \(t\), and let \(A = d!\) be the number of possible ordinal states for embedding dimension \(d\). Under stationarity, let
\[
\pi_i = P(S_t = S_i)
\]
denote the stationary probability of state \(i\), and let
\[
P_{ij} = P(S_{t+1} = S_j \mid S_t = S_i)
\]
denote the one-step transition probability.

The structural entropy of the ordinal-state distribution is
\[
H_q = H(S_t) = -\sum_{i=1}^{A} \pi_i \log \pi_i.
\]
The one-step transition entropy is
\[
H_t = H(S_{t+1} \mid S_t) = -\sum_{i=1}^{A}\sum_{j=1}^{A} \pi_i P_{ij} \log P_{ij}.
\]
The normalized quantities used to construct the \(K_t - K_q\) phase space are
\[
K_q = \frac{H_q}{\log A}, \qquad K_t = \frac{H_t}{\log A}.
\]
The chain rule for entropy gives
\[
H(S_t) + H(S_{t+1} \mid S_t) = H(S_t, S_{t+1})
\]
and the mutual information between consecutive ordinal states is
\[
I(S_t; S_{t+1}) = H(S_t) - H(S_{t+1} \mid S_t).
\]
For a stationary process,
\[
\frac{I(S_t; S_{t+1})}{\log A} = K_q - K_t.
\]
Thus, the vertical coordinate \(K_q\) quantifies the entropy of the symbolic-state distribution, whereas \(K_t\) quantifies the uncertainty remaining in the next state after the present state is known.

\subsection*{S1.2 Bounds and admissible phase-space region}

Because conditional entropy cannot exceed the entropy of the variable being conditioned upon,
\[
H(S_{t+1} \mid S_t) \leq H(S_{t+1}).
\]
Under stationarity, the marginal entropy is time-invariant, so
\[
H(S_t) = H(S_{t+1}) = H_q,
\]
and therefore
\[
0 \leq H_t \leq H_q.
\]
Since the entropy of a discrete variable with \(A\) possible states satisfies
\[
0 \leq H_q \leq \log A,
\]
normalization gives
\[
0 \leq K_t \leq K_q \leq 1.
\]
Consequently, the theoretically admissible \(K_t\)--\(K_q\) phase space is the triangular region bounded by \(K_t = 0\), \(K_t = K_q\), and \(K_q = 1\).

The diagonal
\[
K_t = K_q
\]
corresponds to vanishing mutual information between consecutive ordinal states, since
\[
K_q - K_t = \frac{I(S_t; S_{t+1})}{\log A}.
\]
Points above the diagonal correspond to temporally organized processes for which knowledge of the current ordinal state reduces uncertainty about the next state.

\subsection*{S1.3 Equality conditions}

The condition
\[
K_t = 0
\]
is equivalent to \(H(S_{t+1} \mid S_t) = 0\). Thus, the next ordinal state is completely determined by the current state, up to states of zero probability.

The condition
\[
K_t = K_q
\]
is equivalent to
\[
I(S_t; S_{t+1}) = 0.
\]
Hence, consecutive ordinal states are statistically independent.

Finally,
\[
K_q = 1
\]
requires a uniform stationary distribution,
\[
\pi_i = \frac{1}{A}, \quad i = 1, \ldots, A.
\]
The simultaneous condition
\[
K_t = K_q = 1
\]
therefore corresponds to a uniform state distribution together with statistically independent successive states.

\subsection*{S1.4 Deterministic-transition limit}

Consider a symbolic process in which each occupied state has a unique successor. For every state \(i\) with \(\pi_i > 0\), there exists a state \(j = f(i)\) such that
\[
P_{i,f(i)} = 1,
\]
and all other transition probabilities from state \(i\) are zero.

The conditional entropy of every occupied state is then zero:
\[
-\sum_j P_{ij} \log P_{ij} = 0.
\]
Consequently,
\[
H_t = 0 \quad \text{and} \quad K_t = 0.
\]
Importantly, this does not imply that \(K_q = 0\). A deterministic system may exhibit many ordinal states with a nonuniform or even approximately uniform distribution. Therefore, deterministic temporal organization can occupy a range of \(K_q\) values, while remaining on the \(K_t = 0\) boundary in the ideal symbolic representation.

\subsection*{S1.5 Uniform-transition limit}

Suppose that all ordinal states are equally probable and that the next state is independent of the current state. Then
\[
\pi_i = \frac{1}{A}, \quad i = 1, \ldots, A,
\]
and
\[
P_{ij} = \frac{1}{A}, \quad i,j = 1, \ldots, A.
\]
The state entropy becomes
\[
H_q = -\sum_{i=1}^{A} \pi_i \log \pi_i = \log A.
\]
Similarly,
\[
H_t = -\sum_{i=1}^{A}\sum_{j=1}^{A} \pi_i P_{ij} \log P_{ij} = \log A.
\]
Therefore,
\[
K_q = K_t = 1.
\]
This point represents the limiting case of maximal state diversity together with maximal one-step transition uncertainty.

\subsection*{S1.6 Markov representation and entropy-rate interpretation}

The pair \((\boldsymbol{\pi}, P)\) specifies the stationary one-step symbolic structure, where \(\pi\) is the state distribution and \(P\) is the transition matrix.

For a stationary first-order Markov process,
\[
P(S_{t+1} = j \mid S_t = i, S_{t-1}, \ldots) = P(S_{t+1} = j \mid S_t = i) = P_{ij}.
\]
The stationary distribution satisfies
\[
\pi_j = \sum_{i=1}^{A} \pi_i P_{ij}, \qquad j = 1, \ldots, A.
\]
The one-step conditional entropy is
\[
-\sum_{i=1}^{A}\sum_{j=1}^{A} \pi_i P_{ij} \log P_{ij}.
\]
For a stationary first-order Markov chain, this quantity is also the entropy rate,
\[
H(S_{t+1} \mid S_t).
\]
Thus, \(K_t\) provides a normalized measure of one-step temporal uncertainty and, under the first-order Markov representation, a normalized entropy-rate measure.

\subsection*{S1.7 Finite-window estimation}

For a finite window containing \(W\) observations, let \(N_i\) denote the number of occurrences of ordinal state \(i\). The empirical state probabilities are estimated as
\[
\hat{\pi}_i = \frac{N_i}{W}.
\]
Let \(N_{ij}\) denote the number of observed transitions from state \(i\) to state \(j\). Because a window containing \(W\) observations contains \(W-1\) consecutive transitions, the transition counts satisfy \(\sum_{i,j} N_{ij} = W-1\), whereas the state counts satisfy \(\sum_i N_i = W\). The empirical transition probabilities are
\[
\hat{P}_{ij} = \frac{N_{ij}}{\sum_{j=1}^{A} N_{ij}},
\]
for states with nonzero transition counts.

The finite-window estimates of the two entropy quantities are therefore
\[
\hat{H}_q = -\sum_i \hat{\pi}_i \log \hat{\pi}_i,
\]
and
\[
\hat{H}_t = -\sum_i \hat{\pi}_i \sum_j \hat{P}_{ij} \log \hat{P}_{ij}.
\]
The corresponding normalized estimates are
\[
\hat{K}_q = \frac{\hat{H}_q}{\log A}, \qquad \hat{K}_t = \frac{\hat{H}_t}{\log A}.
\]
All logarithms use the same base, so the normalization removes dependence on that choice.

\subsection*{S1.8 Finite-sample constraint}

Finite-window estimation introduces sampling fluctuations, because the state and transition probabilities are estimated from a limited number of observations.

For a symbolic alphabet containing \(A\) states and a window of length \(W\), the finite-sample resolution depends on both the number of possible states and the number of observations available to estimate their probabilities and transitions.

The criterion used in the present analysis is
\[
K_t K_q > \frac{A-1}{2W}.
\]
This condition provides a practical lower boundary for interpreting estimated phase-space coordinates. Points below this threshold are treated cautiously, because the estimated entropy quantities may be strongly affected by finite-window sampling fluctuations.

The criterion is therefore not an additional dynamical constraint on the exact infinite-sample quantities. Rather, it is an estimation criterion associated with finite windows.

\subsection*{S1.9 Sliding-window trajectories}

To investigate temporal changes in dynamical organization, the time series is divided into overlapping windows of length \(W\). For window \(r\), the corresponding phase-space coordinate is defined as
\[
\mathbf{K}^{(r)} = \left(K_t^{(r)}, K_q^{(r)}\right).
\]
Successive windows are displaced by a fixed step size \(\Delta\). The resulting sequence
\[
\mathbf{K}^{(1)}, \mathbf{K}^{(2)}, \ldots, \mathbf{K}^{(R)}
\]
defines a trajectory through the \(K_t - K_q\) phase space.

The displacement between two consecutive windows is given by
\[
D_r = \sqrt{\left(K_t^{(r+1)} - K_t^{(r)}\right)^2 + \left(K_q^{(r+1)} - K_q^{(r)}\right)^2}.
\]
These quantities characterize the temporal evolution of dynamical organization in the information space independently of the physical units of the original time series.

\subsection*{S1.10 Interpretation of phase-space regions}

The \(K_t - K_q\) plane separates two complementary properties of temporal organization. \(K_q\) describes the diversity and statistical distribution of ordinal states, whereas \(K_t\) describes the uncertainty of transitions between successive states.

Low values of both quantities indicate restricted symbolic-state diversity and relatively predictable temporal evolution. A low \(K_t\) combined with a higher \(K_q\) indicates a system that explores a relatively diverse set of states, while retaining strong temporal organization. Intermediate values of both coordinates can characterize structured nonlinear dynamics, whereas large values of both coordinates indicate increasingly disordered temporal organization.

The diagonal relation
\[
K_t = K_q
\]
marks the limit at which successive states contain no mutual information. Dynamical systems represented below this diagonal retain temporal dependence because
\[
K_q - K_t > 0.
\]
The empirical regions identified from the canonical systems are therefore interpreted comparatively rather than as universal categorical boundaries.

\subsection*{S1.11 Ordinal-state construction}

For a scalar time series \(x_t\), an embedding vector of dimension \(d\) and delay \(\tau\) is defined as
\[
X_t = \left(x_t, x_{t+\tau}, \ldots, x_{t+(d-1)\tau}\right).
\]
The relative ordering of the components of \(X_t\) defines an ordinal pattern. For an embedding dimension \(d\), the number of possible ordinal states is
\[
A = d!.
\]
Ties, when present, require a specified tie-handling convention. In the canonical calculations, the ordinal-state representation is constructed using the same embedding dimension and delay for all series, so that differences in \(K_t\) and \(K_q\) arise from differences in temporal organization rather than from changes in symbolic resolution.

\subsection*{S1.12 Transition matrix and stationary state distribution}

The ordinal-state sequence
\[
S_1, S_2, \ldots, S_M
\]
is used to construct the transition-count matrix
\[
N = (N_{ij}),
\]
where \(N_{ij}\) is the number of observed transitions from state \(i\) to state \(j\).

The corresponding transition matrix satisfies
\[
\sum_j P_{ij} = 1
\]
for every state with nonzero occupancy.

For a stationary process, the state distribution and transition matrix satisfy
\[
\boldsymbol{\pi} P = \boldsymbol{\pi}.
\]
The two quantities entering the \(K_t - K_q\) representation therefore arise from the same symbolic process: \(K_q\) is determined by the stationary state distribution, whereas \(K_t\) is determined by the conditional transition distribution.

\subsection*{S1.13 Relation to mutual information}

The difference between the two normalized coordinates has a direct information-theoretic interpretation:
\[
K_q - K_t = \frac{I(S_t; S_{t+1})}{\log A}.
\]
Thus, the vertical distance of a point from the diagonal is a normalized measure of dependence between successive ordinal states.

A point close to the diagonal has relatively little information about the next state contained in the current state. In contrast, a large separation between \(K_q\) and \(K_t\) indicates stronger temporal dependence.

This relation also shows why \(K_t\) and \(K_q\) should not be interpreted as independent measures. Their difference is itself an information-theoretic quantity describing temporal dependence.

\subsection*{S1.14 Finite-window trajectory resolution}

For sliding-window analysis, the window length \(W\) determines the number of ordinal patterns and transitions available for estimation. Increasing \(W\) generally improves the statistical reliability of estimates but reduces temporal localization.

Conversely, decreasing \(W\) increases temporal resolution, while increasing sampling variability in the estimated state and transition probabilities.

The step size controls the temporal spacing between successive phase-space points. A step smaller than the window length produces overlapping windows and consequently smoother trajectories, whereas a larger step reduces overlap and produces fewer trajectory points.

The selected \(W\) and step therefore represent a compromise between statistical stability and temporal resolution.

\subsection*{S1.15 Canonical-system comparison}

The canonical systems are analyzed using common symbolic and estimation parameters. In the principal analysis,
\[
N = 10{,}000, \quad d = 4, \quad \tau = 1,
\]
so that
\[
A = d! = 24.
\]
For sliding-window analyses, a window length
\[
W = 1000
\]
and step size
\[
\Delta = 1
\]
were used.

For chaotic maps, an initial transient is discarded before the ordinal analysis so that the estimated symbolic statistics represent the developed dynamical regime rather than the chosen initial condition.

The common parameterization allows the positions of the canonical systems in the \(K_t\)--\(K_q\) plane to be compared directly.

\subsection*{S1.16 Interpretation of empirical trajectories}

For empirical time series, a trajectory in the \(K_t\)--\(K_q\) plane represents changes in the temporal organization of the symbolic dynamics over time.

Movement primarily along the \(K_q\) direction indicates changes in the distribution of ordinal states, whereas movement primarily along the \(K_t\) direction indicates changes in one-step transition uncertainty.

A trajectory moving toward larger values of both coordinates indicates increasing state diversity accompanied by increasing transition uncertainty. Movement toward lower \(K_t\) at relatively high \(K_q\) indicates greater temporal organization without a corresponding loss of state diversity.

These interpretations are applied comparatively within each data set and should not be treated as direct physical identification of a unique dynamical mechanism.

\subsection*{S1.17 Scope and limitations of the \(K_t - K_q\) representation}

The \(K_t - K_q\) representation provides an information-theoretic characterization of the ordinal dynamics of a time series. It does not, by itself, identify the underlying governing equations or establish a unique dynamical mechanism.

The coordinates depend on the symbolic representation and therefore on the embedding dimension \(d\), delay \(\tau\), treatment of ties, and available sample size. Sliding-window estimates additionally depend on the window length \(W\) and step size.

Consequently, comparisons between time series should use consistent estimation settings when the objective is to compare their positions in the phase space.

The representation is complementary to other dynamical diagnostics, including Lyapunov exponents, spectral measures, surrogate testing, and conventional statistical descriptors. In particular, a positive Lyapunov exponent and a location in a particular region of the phase space characterize different aspects of temporal organization and should not be regarded as interchangeable measures.

\end{document}